\documentclass[%
 reprint,
superscriptaddress,
nofootinbib,
amsmath,amssymb,
aps,
prl,
]{revtex4-2}

\usepackage{graphicx}
\usepackage{dcolumn}
\usepackage{bm}
\usepackage{blindtext}
\usepackage{xcolor}
\usepackage{amsthm}
\usepackage{amsmath}
\usepackage{amssymb}
\usepackage{relsize}
\usepackage{array}
\usepackage{multirow}
\usepackage{multibib}
\usepackage{hyperref}

\newcommand{\tr}{{\rm tr}}
\newcommand*\pct{\scalebox{.8}{\%}}

\begin{document}

\preprint{DESY 22-}

\title{Novel approach for computing gradients of physical observables}

\author{Simone Bacchio}
\affiliation{
 Computation-based Science and Technology Research Center, The Cyprus Institute, Nicosia, Cyprus
}

\date{\today}

\begin{abstract}
We show that an infinitesimal step of gradient flow can be used for defining a novel approach for computing gradients of physical observables with respect to action parameters. Compared to the commonly used perturbative expansion, this approach does not require calculating any disconnected contribution or vacuum expectation value and can provide results up to three orders of magnitudes more precise. On the other hand, it requires a non-trivial condition to be satisfied by the flow action, the calculation of its force and its Laplacian, and the force of the observable, whose gradient needs to be measured. As a proof of concept, we measure gradients in $\beta$ of Wilson loops in a four-dimensional SU(3) Yang-Mills theory simulated on a $16^4$ lattice using the Wilson action.
\end{abstract}

\maketitle

\section{Introduction} 
In a lattice gauge theory with an action $S_\theta\equiv S(U,\theta)$, the expectation value of an observable $\mathcal{O}(U)$ is given by the path integral
\begin{align}\label{eq:exp_value}
\langle \mathcal{O} \rangle_{\theta} = \frac{1}{\mathcal{Z}_{\theta}} \int \textrm{D}[U] \, \mathcal{O}(U) \, \exp(-S(U,\theta))\,,
\end{align}
where 
$\mathcal{Z}_{\theta}\equiv\int \textrm{D}[U] \,\exp(-S(U,\theta))$ is the partition function and $\theta$ is a parameter of the action.
The focus of this work is the calculation of the gradient of the expectation value with respect to the parameter $\theta$, which might result in a challenging task since the so-called sea effects have to be measured.
Indeed, it is standard to perturbatively expand the expectation value with respect to an infinitesimal change of the parameter, $d\theta$, obtaining
\begin{align*}
\langle \mathcal{O}_{\theta + d\theta} \rangle_{\theta+ d\theta} = \frac{\langle d\theta \frac{\partial \mathcal{O}_\theta}{\partial\theta} +  \mathcal{O}_\theta \exp(-d\theta\frac{\partial S_\theta}{\partial\theta})\rangle_\theta}{\langle\exp(-d\theta\frac{\partial S_\theta}{\partial\theta})\rangle_\theta}+O(d\theta^2)
\end{align*}
and, thus, the gradient with respect to $\theta$ is given by
\begin{equation}\begin{aligned}\label{eq:standard_FH_1}
\frac{d\langle \mathcal{O}_\theta \rangle_\theta}{d\theta} &= \lim_{d\theta\rightarrow 0}\frac{\langle \mathcal{O}_{\theta + d\theta} \rangle_{\theta+ d\theta} - \langle \mathcal{O}_{\theta} \rangle_{\theta}}{d\theta}\\
&= \langle \frac{\partial \mathcal{O}_\theta}{\partial\theta} -\mathcal{O}_\theta\frac{\partial S_\theta}{\partial\theta}\rangle_\theta + \langle\mathcal{O}_\theta\rangle_\theta\langle\frac{\partial S_\theta}{\partial\theta}\rangle_\theta\,,
\end{aligned}\end{equation}
where, in the above equations, for generality's sake, we have introduced an optional explicit dependence on $\theta$ also in the observable  $\mathcal{O}_\theta\equiv \mathcal{O}(U,\theta)$. Since on both sides of Eq.~\eqref{eq:standard_FH_1} all terms depend on $\theta$, we rewrite the equation with the shorthand notation 
\begin{equation}\begin{aligned}\label{eq:standard_FH}
\frac{d\langle \mathcal{O} \rangle}{d\theta} &= \langle \frac{\partial \mathcal{O}}{\partial\theta} -\mathcal{O}\frac{\partial S}{\partial\theta}\rangle + \langle\mathcal{O}\rangle\langle\frac{\partial S}{\partial\theta}\rangle\,
\end{aligned}\end{equation}
and we refer to
\begin{align*}
     \langle \frac{\partial \mathcal{O}}{\partial\theta}\rangle\,:&\quad\text{as connected contribution,}\\
     \langle\mathcal{O}\frac{\partial S}{\partial\theta}\rangle\,:&\quad\text{as disconnected contribution, and}\\ 
     \langle\mathcal{O}\rangle\langle\frac{\partial S}{\partial\theta}\rangle\,:&\quad\text{as vacuum expectation value.}
\end{align*}
Disconnected contributions are well-known to be noisy.
Furthermore, when the vacuum expectation value is non-zero, the latter and the disconnected contributions are usually large values, whose difference needs to be taken accurately for reliably measuring the gradient. 

In this work, we present a novel and alternative approach, based on gradient-flow techniques~\cite{luscher2010trivializing}, which is free from any of the aforementioned problems. Namely, if a flow action that flows along the parameter $\theta$ is found, then gradients of observables can be computed directly using the ODE of the operator. Our numerical results show that sea effects can be computed up to three orders of magnitude more precisely than the perturbative expansion in Eq.~\eqref{eq:standard_FH},  when an exact flow action is found. Therefore, this approach can possibly be used to improve the accuracy of gradients in, for example, among others: \\
i.) applications of the Feynman-Hellmann theorem~\cite{HORSLEY2012312,CSSM:2014uyt,Chambers:2015bka,Bouchard:2016heu,Chang:2018uxx,Can:2020sxc},\\ ii.) leading isospin breaking corrections~\cite{deDivitiis:2011eh,deDivitiis:2013xla,BMW:2014pzb,Giusti:2017dmp,Borsanyi:2020mff}, \\
iii.) leading QED corrections~\cite{Giusti:2017dwk,DiCarlo:2019thl,Borsanyi:2020mff}, \\
iv.) fine-tuning of simulation parameters~\cite{ExtendedTwistedMass:2022jpw}, \\
v.) leading contribution of the QCD $\Theta$-term to the neutron electric dipole moment~\cite{Dragos:2019oxn,Alexandrou:2020mds,Bhattacharya:2021lol}.

In the following, we present the novel approach in the form of a theorem, followed by its proof and a numerical case study.

\section{Main result}
\paragraph{\textbf{Notation.}} In the following,\\
\textbullet~~$\tilde{S}$ is referred to as \textit{flow action} and is to-be-determined;\\
\textbullet~~$\mathcal{L}_0=-\sum_{x,\mu,a}\partial_{x,\mu}^a\partial_{x,\mu}^a$ denotes the Laplacian;\\
\textbullet~~$(A,B)=\sum_{x,\mu,a} A^a_\mu(x)B^a_\mu(x)$ denotes the scalar product over algebra-valued fields; and\\
\textbullet~~$\partial$ is the force of a scalar function, defined as
\begin{align}
    \partial^a_{x, \mu} f(U) = \left. \frac{d}{d \tau} f(U_\tau) \right|_{\tau = 0}
    \label{eq:def_group_deriv}
\end{align}
with 
\begin{align}
     U_\tau (y, \nu) = \begin{cases}
    e^{\tau T^a} U(x, \mu), & \text{for } (x, \mu) = (y, \nu) \,, \\
    U(y, \nu) &  \text{for } (x, \mu) \neq (y, \nu)  \,.
  \end{cases}
\end{align}

\paragraph{\textbf{Theorem, gradient-flow approach.}}
The gradient of an expectation value with respect to a parameter $\theta$ of the action is given by
\begin{align}\label{eq:new_FH}
\frac{d\langle \mathcal{O} \rangle}{d\theta} 
&= \langle \frac{\partial \mathcal{O}}{\partial\theta} +\left(\partial\mathcal{O},\partial \tilde{S}\right)-\mathcal{O}\mathcal{C}\rangle + \langle\mathcal{O}\rangle\langle \mathcal{C}\rangle\,,
\end{align}
where
\begin{align}\label{eq:cost_fucntion}
    \mathcal{C} = \mathcal{L}_0 \tilde{S} +\left(\partial S, \partial \tilde{S} \right)  +\frac{\partial S}{\partial\theta}\,.
\end{align}

\paragraph{\textbf{Corollary, ideal gradient-flow approach.}}
If the flow action $\tilde{S}$ is such that $ \mathcal{C}=$~constant,
then Eq.~\eqref{eq:new_FH} simplifies to 
\begin{align}\label{eq:simple_FH}
\frac{d\langle \mathcal{O} \rangle}{d\theta} &= \langle \frac{\partial \mathcal{O}}{\partial\theta} +\left(\partial\mathcal{O},\partial \tilde{S}\right)\rangle\,.
\end{align}
We, therefore, refer as \textit{ideal} flow action to an $\tilde{S}$ such that
\begin{align}\label{eq:ideal_S_tilde}
    \mathcal{L}_0 \tilde{S} +\left(\partial S, \partial \tilde{S} \right)  +\frac{\partial S}{\partial\theta}={\rm constant}\,,
\end{align}

\paragraph{\textbf{Corollary, standard perturbative approach.}}
If $\tilde{S}\equiv$ constant, then Eq.~\eqref{eq:new_FH} simplifies to Eq.~\eqref{eq:standard_FH}, since  
$ \mathcal{L}_0 \tilde{S} = \partial\tilde{S}=0$ and $\mathcal{C} = \partial S_\theta/\partial\theta$. \\

\paragraph{\textbf{Summary.}}
The corollaries are two special applications of the theorem and our numerical results show that gradients computed using the ideal gradient-flow approach are significantly more precise than those computed using the standard perturbative approach. If Eq.~\eqref{eq:ideal_S_tilde} is not satisfied exactly, then the theorem can be used to obtain correct and possibly improved results.

\section{Proof} 
The proof is based on gradient-flow techniques that have been introduced in Ref.~\cite{luscher2010trivializing}, which we refer to for further details.
Under a change of variables $U=\mathcal{F}(V)$, the expectation value of an observable transforms as
\begin{equation}\begin{aligned}\label{eq:exp_value_flow}
    \langle \mathcal{O} \rangle &= \frac{1}{\mathcal{Z}} \int \textrm{D}[U] \, \mathcal{O}(U) \, \exp(-S(U))\\
    &= \frac{1}{\mathcal{Z}_\mathcal{F}} \int \textrm{D}[V] \, \mathcal{O}(\mathcal{F}(V)) \, \exp(-S_\mathcal{F}(V))\,
\end{aligned}\end{equation}
where $\mathcal{Z}_\mathcal{F} \equiv \int \textrm{D}[V] \exp(-S_\mathcal{F}(V))$ and
\begin{equation}\begin{aligned}\label{eq:base_action}
    S_{\mathcal{F}}(V)&=S(\mathcal{F}(V)) -\ln \det \mathcal{F}_* (V)
\end{aligned}\end{equation}
with $\mathcal{F}_*$ being the Jacobian of the transformation. In the following, we consider an infinitesimal transformation of the field employing an Euler integration step and the force of an action $\tilde{S}$
\begin{equation}
    \begin{aligned}
    \mathcal{F}(V) &= \exp(-\epsilon\,\partial \tilde{S}(V)) V\\
    &= V -\epsilon\,\partial \tilde{S}(V) V + O(\epsilon^2)
    \end{aligned}
    \quad\text{with}\quad0\!<\!\epsilon\!\ll\!1.
\end{equation}
Two properties of such flow are~\cite{luscher2010trivializing}
\begin{equation}\label{eq:logdet}
\ln \det \mathcal{F}_* (V) = \epsilon \, \mathcal{L}_0 \tilde{S}(V) + O(\epsilon^2)
\end{equation}
and
\begin{equation}\label{eq:O(F(U))}
\mathcal{O}(\mathcal{F}(V)) = \mathcal{O}(V) - \epsilon \, \left(\partial\mathcal{O}(V),\partial \tilde{S}(V)\right) + O(\epsilon^2)\,.
\end{equation}
By using Eqs.~\eqref{eq:logdet} and~\eqref{eq:O(F(U))} in Eq.~\eqref{eq:base_action}, we obtain
\begin{align}\label{eq:infinitesimal_flow}
    S_\mathcal{F} = S -\epsilon \left[ \left(\partial S, \partial \tilde{S} \right) + \mathcal{L}_0 \tilde{S}\right]+O(\epsilon^2),
\end{align}
where all terms now depend on the field $V$. 
The intent is then to interpret the difference $S_\mathcal{F}-S$ as an infinitesimal change of action along the parameter $\theta$, requiring
\begin{align}
    \frac{\partial S}{\partial\theta} = \lim_{\epsilon\rightarrow 0}\frac{S_\mathcal{F}-S}{\epsilon} = -\left(\partial S, \partial \tilde{S} \right) - \mathcal{L}_0 \tilde{S}\,.
\end{align}
We then define $\mathcal{C}$, in Eq.~\eqref{eq:cost_fucntion}, as the difference between the left- and right-hand sides. By using $\mathcal{C}$ in Eq.~\eqref{eq:infinitesimal_flow}, we can remove any dependence on $\tilde{S}$ obtaining
\begin{align}\label{eq:final_S0}
    S_\mathcal{F} = S + \epsilon\left(\frac{\partial S}{\partial\theta}-\mathcal{C}\right)+O(\epsilon^2)\,.
\end{align}
In the following, we track explicitly the dependence on the parameter $\theta$, and from Eq.~\eqref{eq:final_S0} we obtain
\begin{align}\label{eq:final_S0_2}
    S_\mathcal{F} = S_{\theta+\epsilon} -\epsilon\,\mathcal{C_\theta}+O(\epsilon^2)\,.
\end{align}
By using the latter in the definition of $\mathcal{Z}_\mathcal{F}$, we obtain
\begin{align}
    \mathcal{Z}_\mathcal{F} = \mathcal{Z}_{\theta+\epsilon}-\epsilon\,\mathcal{Z}_{\theta}\langle\mathcal{C_\theta}\rangle_\theta+O(\epsilon^2)\,.
\end{align}
Finally, by using the above in Eq.~\eqref{eq:exp_value_flow}, we obtain
\begin{equation}\begin{aligned}\label{eq:exp_value_flow_e}
    \langle \mathcal{O}_\theta \rangle_\theta = \langle \mathcal{O}_\theta \rangle_{\theta+\epsilon}&+\epsilon\,\langle\mathcal{O}_\theta\mathcal{C}_\theta\rangle_{\theta}-\epsilon\,\langle\mathcal{O}_\theta\rangle_{\theta}\langle \mathcal{C}_{\theta}\rangle_{\theta}\\
    &-\epsilon\,\langle\left(\partial\mathcal{O}_\theta,\partial \tilde{S}_\theta\right)\rangle_\theta+O(\epsilon^2)
\end{aligned}\end{equation}
and the theorem is proved since
\begin{align}
\frac{d\langle \mathcal{O}_\theta \rangle_\theta}{d\theta} &= \lim_{\epsilon\rightarrow 0}\frac{\langle \mathcal{O}_{\theta + \epsilon} \rangle_{\theta+ \epsilon} - \langle \mathcal{O}_{\theta} \rangle_{\theta}}{\epsilon}\\
&=\langle \frac{\partial \mathcal{O}_\theta}{\partial\theta} +\left(\partial\mathcal{O}_\theta,\partial \tilde{S}_\theta\right)-\mathcal{O}_\theta\mathcal{C}_\theta\rangle_\theta + \langle\mathcal{O}_\theta\rangle_\theta\langle \mathcal{C}_\theta\rangle_\theta\,.\nonumber
\end{align}

\section{Case study and numerical results}

\begin{figure}
    \centering
    \includegraphics[width=\linewidth]{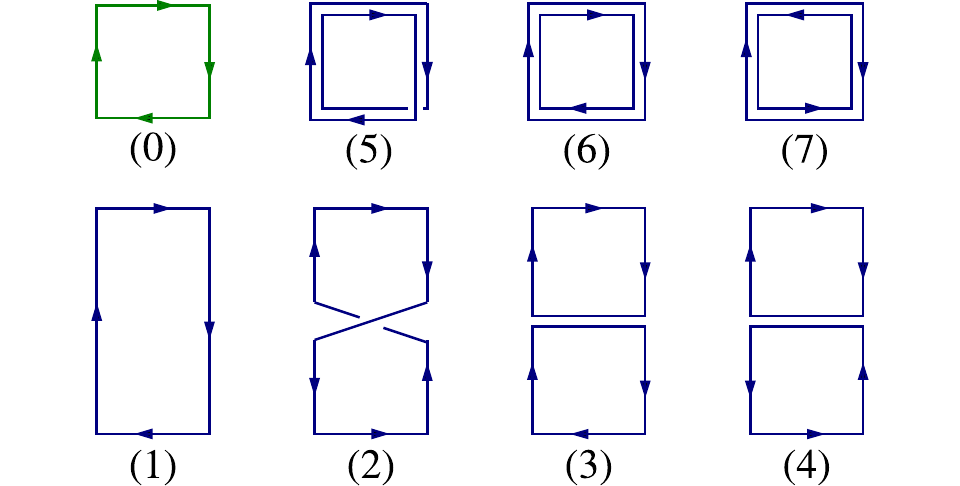}
    \caption{Representation of the loops entering the $\beta$-expansion up to the NLO. The  plaquette, shown by the green loop, enters the LO expansion, while the loops shown in blue enter the NLO. The enumeration of the loops follows Ref.~\cite{luscher2010trivializing}.}
    \label{fig:loops}
\end{figure}

\paragraph{\textbf{Wilson action.}} To demonstrate the effectiveness of Eq.~\eqref{eq:new_FH} compared to Eq.~\eqref{eq:standard_FH}, we consider a four-dimensional SU(3) Yang-Mills theory defined on a lattice using the standard Wilson action
\begin{equation}\label{eq:wilson_action}
    S_W(\beta, U) = -\frac{\beta}{6} \mathcal{W}_0(U),
\end{equation}
where $\mathcal{W}_0$ denotes the sum of plaquettes. We aim at computing the slope in $\beta$ of various observables and, therefore, the approach outlined here requires a flow action $\tilde{S}$ that satisfies Eq.~\eqref{eq:ideal_S_tilde} in the following way,
\begin{equation}\label{eq:target_wilson}
    \mathcal{L}_0 \tilde{S}-\frac{\beta}{6}\left(\partial \mathcal{W}_0,\partial\tilde{S}\right)-\frac{1}{6} \mathcal{W}_0 = \text{ constant}\,.
\end{equation}
Inspired by L\"uscher's $t$-expansion~\cite{luscher2010trivializing}, it is easy to show that an analytical solution to Eq.~\eqref{eq:target_wilson} is
\begin{equation}\begin{aligned}\label{eq:beta_expansion}
    \tilde{S} &= \frac{1}{6}\sum_{k=0}^{\infty} \left(\frac{\beta}{6}\right)^k\tilde{S}^{(k)} \qquad~~\,\text{with}\\
    \tilde{S}^{(0)} &= \mathcal{L}_0^{-1}\mathcal{W}_0 = \frac{3}{16} \mathcal{W}_0 \quad\quad~~~{\rm and}\\
    \tilde{S}^{(k)} &= \mathcal{L}_0^{-1}\left(\partial \mathcal{W}_0,\partial \tilde{S}^{(k-1)}\right)\quad{\rm for~}k>0\,,
\end{aligned}\end{equation}
which we refer to as $\beta$-expansion and it is equivalent to L\"uscher's $t$-expansion evaluated at $t=\beta$ divided by $\beta$.
We refer to $\tilde{S}^{(0)}$ as the leading order (LO) and to $\tilde{S}^{(1)}$ as the next-to-leading order (NLO). The calculation of the NLO can be found in Ref.~\cite{luscher2010trivializing}, where the term 
$\left(\partial \mathcal{W}_0,\partial \mathcal{W}_0\right)$ is computed. It results in a linear combination of the Wilson loops depicted in Fig.~\ref{fig:loops}, which are defined as 
\begin{equation}\begin{aligned}\label{eq:loops}
\mathcal{W}_i &= \sum_{C\in\Gamma_i} \tr\{U(C)\} &{\rm for~} i=0,1,2,5\\
\mathcal{W}_i &=\!\!\! \sum_{C,C'\in\Gamma_i}\!\!\! \tr\{U(C)\} \tr\{U(C')\} &{\rm for~} i=3,4,6,7
\end{aligned}\end{equation}
where $\Gamma_i$ are all unique loops for a given shape, including loops in the perpendicular direction -- i.e. chair-shaped -- for $i=$1,2,3,4. \\

\begin{figure}
    \centering
    \includegraphics[width=\linewidth]{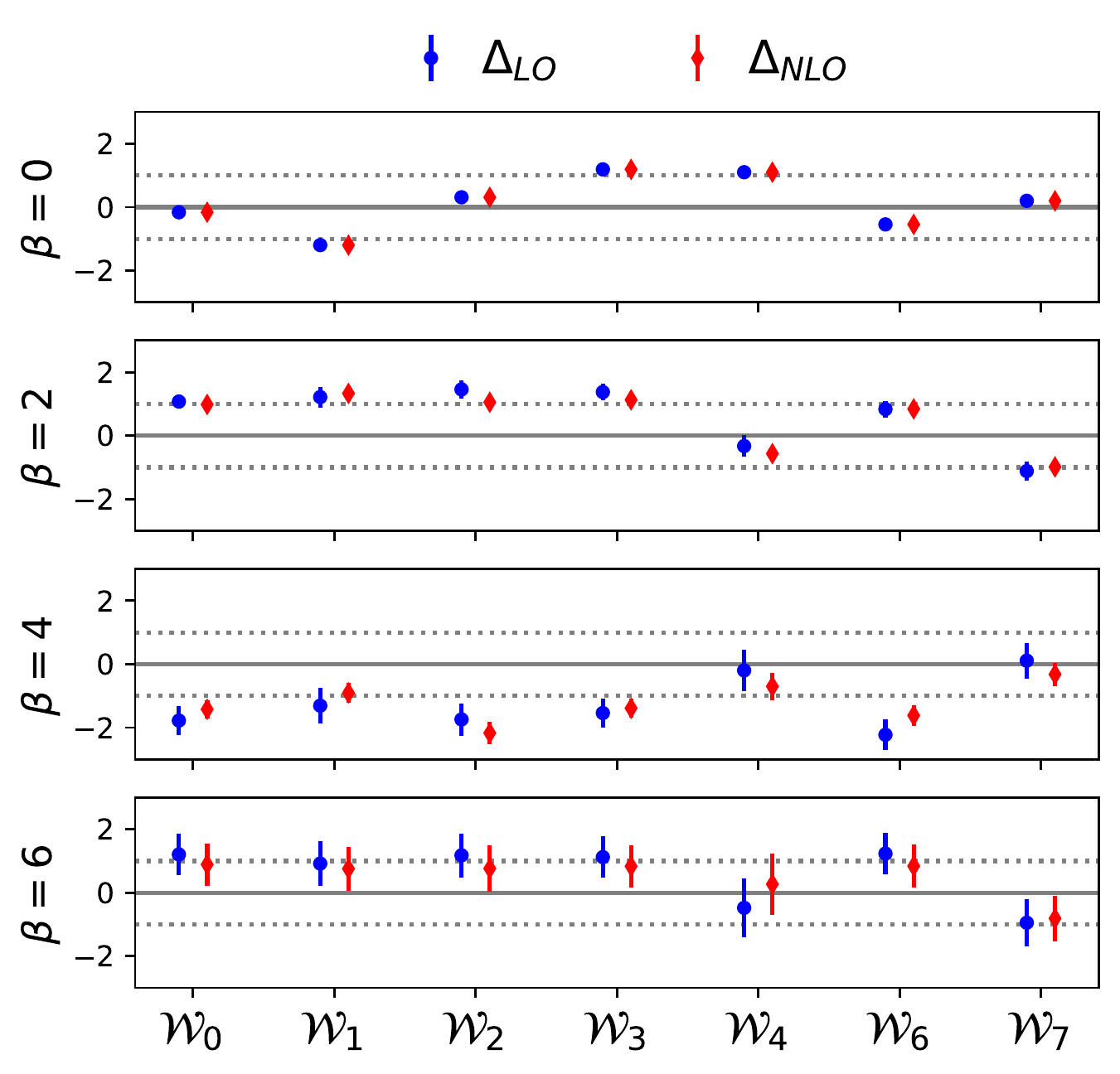}
    \caption{Accuracy on the gradients in $\beta$ for the Wilson loops, $\mathcal{W}_i$ depicted in Fig.~\ref{fig:loops}, measured using the perturbative expansion (PE) in Eq.~\eqref{eq:standard_FH} or alternatively using the new approach in Eq.~\eqref{eq:new_FH} with the LO or the NLO terms for the flow action $\tilde{S}$ in Eq.~\eqref{eq:beta_expansion}. We depict the deviation $\Delta_T$ from the values measured using PE normalized such that the errors of PE are unitary, see Eq.~\eqref{eq:delta_T}. Thus, the continuous lines are the central values of PE, while the dashed lines are plus or minus one unit of the error of PE. Results at various $\beta$ are computed over ten thousand configurations with lattice size $16^4$, simulated with HMC and separated by 4 MDUs.}
    \label{fig:results}
\end{figure}

\begin{table}
    \centering
    \renewcommand{\arraystretch}{1.2}
    \begin{tabular}{rr|r|r|r|r|r|r|r|}
        \cline{3-9}
          && \,$\beta=0$ & \,$\beta=1$ & \,$\beta=2$ & \,$\beta=3$ & \,$\beta=4$ & \,$\beta=5$ & \,$\beta=6$ \\
         \hline
         \multirow{2}{*}{\!\!LO~~}&min~ & 0.08\pct & 12.0\pct & 23.1\pct & 37.6\pct & 45.4\pct & 67.6\pct & 65.3\pct \\
         &max~ & 0.45\pct & 16.7\pct & 33.5\pct & 48.2\pct & 66.2\pct & 81.1\pct & 92.5\pct \\
         &$\!\!\!\!\!\!\sigma_C/\sigma_{S'}$ & 0.00\pct & 15.0\pct & 29.2\pct & 47.1\pct  & 65.6\pct & 75.6\pct & 102.0\pct \\
         \hline
         \multirow{2}{*}{\!\!NLO}&min~ & 0.08\pct & 1.8\pct & 6.9\pct & 17.0\pct & 29.6\pct & 55.3\pct & 68.9\pct \\
         &max~ & 0.45\pct & 2.5\pct & 10.4\pct & 22.4\pct & 43.9\pct & 66.3\pct & 99.1\pct \\
         &$\!\!\!\!\!\!\sigma_C/\sigma_{S'}$ & 0.00\pct & 2.4\pct & 9.7\pct & 22.8\pct  & 42.3\pct & 62.2\pct & 100.7\pct \\\hline\hline
    \end{tabular}
    \renewcommand{\arraystretch}{1}
    \caption{Smallest and largest value for the ratio of the error obtained using the new approach defined by Eq.~\eqref{eq:new_FH} and the error of the perturbative expansion given by  Eq.~\eqref{eq:standard_FH} over the measured Wilson loops. Results are given for the cases when the LO or the NLO is used in the flow action. We report also the ratio between the standard deviation of $C$ and the one of $S'=\partial S/\partial\beta$, noting a strong connection between this ratio and the one between the errors. }
    \label{tab:errors_ratio}
\end{table}

\paragraph{\textbf{Gradients of Wilson loops.}}
Since the convergence of the $\beta$-expansion in Eq.~\eqref{eq:beta_expansion} deteriorates as $\beta$ grows, we show results for simulations performed at various values of $\beta$ from $\beta=0$ up to $\beta=6$. At $\beta=0$ the LO of the $\beta$-expansion solves Eq.~\eqref{eq:target_wilson} exactly, while, on the other hand, the region of physical interest starts from $\beta\gtrsim 5.8$, where a lattice spacing of about $0.14$~fm is measured~\cite{Schaefer:2010hu}.
We compute the slope in $\beta$ for a set of Wilson loops, $\mathcal{W}_i$, using either the perturbative expansion (PE) or the new approach using the LO or the NLO of Eq.~\eqref{eq:beta_expansion} computed at each $\beta$. For all experiments, we use a lattice of size $16^4$ and measure expectation values over ten thousand configurations simulated at each value of $\beta$ using an HMC algorithm. To compare gradients for various Wilson loops, we study the deviation of the value measured using an approach ``$T$'' from the weighted average of all approaches, namely
\begin{equation}\label{eq:delta_T}
    \Delta_{T,i} = \frac{1}{\sigma_{\rm PE, i}}\left(\frac{d\langle W_i\rangle}{d\beta}\Bigg|_T - \frac{d\langle W_i\rangle}{d\beta}\Bigg|_{\rm PE}\right)\,,
\end{equation}
where  $\frac{d\langle W_i\rangle}{d\beta}\big|_T$ is the value from the approach $T$ and we normalize the results using the error from the perturbative expansion $\rm PE$, such that the error of $\Delta_{{\rm PE},i}$ is always unity.
Results are depicted in Fig.~\ref{fig:results} and summarized in Table~\ref{tab:errors_ratio}. As can be seen, we observe excellent statistical agreement between the results measured using the perturbative expansion and the novel approach discussed in this work, supporting the correctness of the derivation. We also observe an impressive improvement in the error achieved at small $\beta$. This is highlighted in Table~\ref{tab:errors_ratio}, in which we list the minimum and maximum values of the ratio between the errors obtained using the various approaches. At $\beta=0$, results for LO and NLO are the same, since LO solves exactly Eq.~\eqref{eq:target_wilson}. Here we achieve the largest improvement for the signal-to-noise ratio, which is of more than three orders of magnitude in calculating the slope in $\beta$ of the plaquette. Additionally, looking at Table~\ref{tab:errors_ratio}, we observe a very strong connection between the improvement in the errors and the ratio of the standard deviations of $C$ and $\partial S/\partial\beta$. The latter, therefore, can be used to have an estimate of the gain one would achieve in a generic situation.
\\

\begin{figure}
    \centering
    \includegraphics[width=\linewidth]{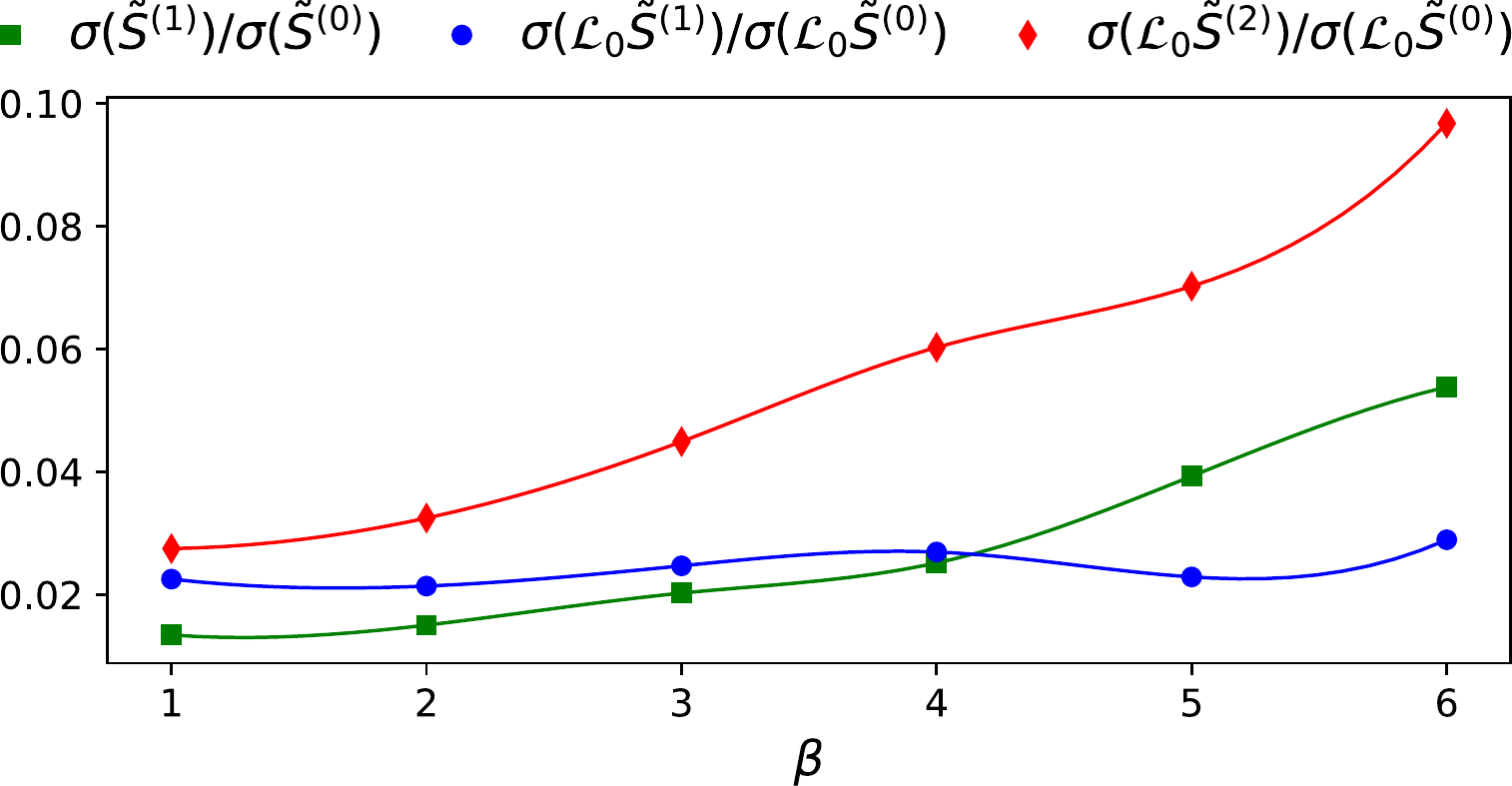}
    \caption{Study of the convergence of the $\beta$-expansion, showing ratios between the standard deviation of the LO, the NLO, and the next-to-next-leading-order (NNLO)  of the flow action or its Laplacian. Lines are interpolations for guiding the eye.}
    \label{fig:convergence}
\end{figure}

\paragraph{\textbf{Convergence of the $\beta$-expansion.}} On a side note, our experiments let us comment on the convergence of the $\beta$-expansion in Eq.~\eqref{eq:beta_expansion}. This provides some interesting insights into the ability to flow in $\beta$ using gradient flows~\cite{Bacchio:2022vje}. On one side, if all $\tilde{S}^{(k)}$ have similar magnitude, the series would be converging only for $\beta<6$. On the other hand, $\tilde{S}^{(k)}$ might decrease in magnitude at increasing $k$, therefore converging for all $\beta$. In Ref.~\cite{luscher2010trivializing}, an analytical study on the convergence of the $t$-expansion is found and similar arguments can be applied to the $\beta$-expansion since they share the same origin. In the following, on the other hand, we provide numerical evidence for the first few orders. In our numerical experiments we have computed the actions $\tilde{S}^{(0)}$ and $\tilde{S}^{(1)}$. Additionally, we can measure the value of $\mathcal{L}_0\tilde{S}^{(2)}$ from its definition, namely
\begin{equation}\label{eq:nnlo}
    \mathcal{L}_0\tilde{S}^{(2)} = \left(\partial\mathcal{W}_0, \partial\tilde{S}^{(1)}\right)\,,
\end{equation}
which is the scalar product of the forces of $\mathcal{W}_0$ and of $\tilde{S}^{(1)}$, both computed in our experiments. Results for the ratio of the first three orders are given in Fig.~\ref{fig:convergence}, in which we analyze the standard deviation of these terms and not their value since the flow action is defined up to a constant~\eqref{eq:target_wilson}. While $\sigma(\tilde{S}^{(1)})$ is almost two orders of magnitude smaller than $\sigma(\tilde{S}^{(0)})$, the ratio $\sigma(\mathcal{L}_0\tilde{S}^{(2)})/\sigma(\mathcal{L}_0\tilde{S}^{(0)})$ is larger than $\sigma(\mathcal{L}_0\tilde{S}^{(1)})/\sigma(\mathcal{L}_0\tilde{S}^{(0)})$ and increases in $\beta$, pointing in the direction that the various orders might have similar magnitude. If this is true, then the $\beta$-expansion is converging only for $\beta<6$ while at larger $\beta$ possibly all terms contribute to the flow action, making it an extensive operator in the region where the regime of physical interest begins. This is unfortunate for flow-based trivializing maps since flowing at large $\beta$ becomes very challenging. Indeed, to date, only flowing at very small $\beta$ has been proven successful in four dimensions~\cite{Abbott:2023thq}.

\section{Conclusions}

In this work, we have presented a novel approach for computing gradients of observables with respect to action parameters, which, in its simpler form given in  Eq.~\eqref{eq:simple_FH}, is free of any disconnected contribution or vacuum expectation value. We provide numerical evidence within our case study that when a flow action solution to Eq.~\eqref{eq:ideal_S_tilde} is found, the new approach provides a significant improvement in the signal-to-noise ratio compared to the commonly-used perturbative expansion. We use, as a case study, the calculation of the slope in $\beta$ of Wilson loops in pure-gauge SU(3) Yang-Mills theories. In this situation, we are able to provide in Eq.~\eqref{eq:beta_expansion} an analytic solution to the required flow action in terms of a series expansion, referred to as $\beta$-expansion. We compute the first two orders of the series and show results at various $\beta$ for many Wilson loops. At $\beta=0$, the expansion converges at the leading order, and in this case, we obtain the largest improvement in the errors. At finite $\beta$, instead, the expansion converges in orders of $\beta/6$. Therefore, a reduction in the errors is seen only for $\beta<6$, which is a region of non-physical interest. On the other hand, these results are only meant as proof of concept, since the approach will change according to the application. Indeed, each case should be studied independently, searching for an appropriate flow action, that satisfies Eq.~\eqref{eq:ideal_S_tilde} for a desired action $S$ and parameter of the action $\theta$. Where possible, this can be done analytically or, alternatively, it can be done numerically using a parametric definition of $\tilde{S}$~\cite{Bacchio:2022vje} and machine learning techniques to find the optimal flow action. In this case, for the training, we suggest minimizing either the variance of $\langle \mathcal{C}\rangle$, see Eq.~\eqref{eq:cost_fucntion}, since the final objective is to make it constant, i.e.~Eq.~\eqref{eq:ideal_S_tilde}; or to perform an observable-dependent training by minimizing the value or the variance of $\langle \mathcal{O}\rangle\langle \mathcal{C}\rangle - \langle \mathcal{O}\mathcal{C}\rangle$. We have tested such an approach for the case study presented here and we have obtained results compatible with the analytical solution and no further improvement was observed. On the other hand, for all other cases where an analytical solution is not easy to find, this is a new valuable application of machine-learning techniques. \\

\begin{acknowledgments}
 \paragraph{\textbf{Acknowledgments.}} The author wishes to thank Pan Kessel, Stefan Schaefer, and Lorenz Vaitl for useful discussions and the enjoyable collaboration on gradient flows. The author acknowledges financial support from the Cyprus Research and Innovation Foundation under the project QC4LGT (grant agreement No.~EXCELLENCE/0421/0019) and 3D-NUCLEON (grant agreement No.~EXCELLENCE/0421/0043).
Numerical results were obtained using the Cyclamen cluster of The Cyprus Institute equipped with P100 GPUs. The software is implemented using the Lyncs-API~\cite{Bacchio:2022bjk}, its interface to QUDA~\cite{Yamamoto:2022ygt} and the QUDA library~\cite{Clark:2009wm}. All figures are available under a CC BY 4.0 license\footnote{https://creativecommons.org/licenses/by/4.0/}.
\end{acknowledgments}

\bibliography{main}
\nopagebreak
\appendix
\include{supplemental}
\end{document}